\documentclass[%
reprint,
superscriptaddress,
amsmath,
amssymb,
aps,
prl,
longbibliography
]{revtex4-1}

\usepackage{times}

\usepackage{graphicx}
\usepackage{empheq}
\usepackage{amsmath}
\usepackage{amsfonts}
\usepackage{dsfont}
\usepackage{amssymb}
\usepackage{amsthm}
\usepackage{mathtools}
\usepackage{float}
\usepackage{color}
\usepackage{placeins}

\newcommand{\ket}[1]{|{#1}\rangle}
\newcommand{\bra}[1]{\langle{#1}|}

\newcommand{\beq}{\begin{equation}}
\newcommand{\eeq}{\end{equation}}

\begin{document}

\author{Pengcheng Yang}
\affiliation{School of Physics, International Joint Laboratory on Quantum Sensing and Quantum Metrology, Huazhong University of Science and Technology, Wuhan 430074, China}
\author{Min Yu}
\affiliation{School of Physics, International Joint Laboratory on Quantum Sensing and Quantum Metrology, Huazhong University of Science and Technology, Wuhan 430074, China}
\author{Ralf Betzholz} 
\email{ralf\_betzholz@hust.edu.cn}
\affiliation{School of Physics, International Joint Laboratory on Quantum Sensing and Quantum Metrology, Huazhong University of Science and Technology, Wuhan 430074, China}
\author{Christian Arenz} 
\affiliation{Frick Laboratory, Princeton University, Princeton New Jersey 08544, USA}
\author{Jianming Cai}
\email{jianmingcai@hust.edu.cn}
\affiliation{School of Physics, International Joint Laboratory on Quantum Sensing and Quantum Metrology, Huazhong University of Science and Technology, Wuhan 430074, China}

\title{Complete Quantum-State Tomography with a Local Random Field}
\date{\today}

\begin{abstract}
Single-qubit measurements are typically insufficient for inferring arbitrary quantum states of a multiqubit system. We show that if the system can be fully controlled by driving a single qubit, then utilizing a local random pulse is almost always sufficient for complete quantum-state tomography. Experimental demonstrations of this principle are presented using a nitrogen-vacancy (NV) center in diamond coupled to a nuclear spin, which is not directly accessible. We report the reconstruction of a highly entangled state between the electron and nuclear spin with fidelity above 95\% by randomly driving and measuring the NV-center electron spin only. Beyond quantum-state tomography, we outline how this principle can be leveraged to characterize and control quantum processes in cases where the system model is not known.     
\end{abstract}

\maketitle
\textit{Introduction.--} The ability to infer the full state of a quantum system is crucial for benchmarking and controlling emerging quantum technologies. In theory, this task can be accomplished by measuring an \emph{informationally complete}~\cite{busch1991informationally} set of observables, whose corresponding expectation values allow to reconstruct the quantum state of the system. In practice, measuring observables that are informationally complete typically requires access to each system component. While compressed sensing techniques can significantly improve the efficiency of reconstructing low-rank quantum states~\cite{candes2011probabilistic,gross2010quantum, flammia2012quantum,ohliger2013efficient, kalev2015quantum, shabani2011efficient, christandl2012reliable,riofrio2017experimental, steffens2017experimentally, cramer2010efficient, kalev2015quantum}, the problem of identifying an arbitrary state of a complex quantum system with limited measurement access (e.g., to a single qubit only) remains \cite{merkel2010random, smith2013quantum,Chantasri2019}. For example, one task of practical importance in the development of solid-state quantum devices~\cite{cai2013large,bradley2019, zhao2012sensing,Kolkowitz2012,Taminiau2012} is the complete characterization of coupled spin states. However, when nuclear spins are involved, access to the full system is limited due to their small magnetic moment. Even in settings where full access is currently possible (e.g., proof-of-principle few-qubit devices), this requirement becomes daunting as the complexity of the system (e.g., the number of qubits) grows.

A typical strategy for addressing these challenges   is to create otherwise inaccessible observables. This can be accomplished by (i) deterministically applying unitary operations that transform an accessible observable into the desired inaccessible ones~\cite{silberfarb2005quantum, deutsch2010quantum, merkel2010random,liu2019pulsed}, typically via properly tailored classical fields, or (ii) randomly creating an informationally complete set of observables by approximating random unitary transformations through so-called unitary $t$-designs~\cite{flammia2012quantum, ohliger2013efficient}. However, both of these procedures can be highly demanding. While (i) does not require full system access, it does require identifying  and accurately implementing the necessary classical fields; (ii), on the other hand, can be carried out with elementary gate operations, but typically necessitates full system access.

\begin{figure}[t]
\includegraphics[width=0.88\linewidth]{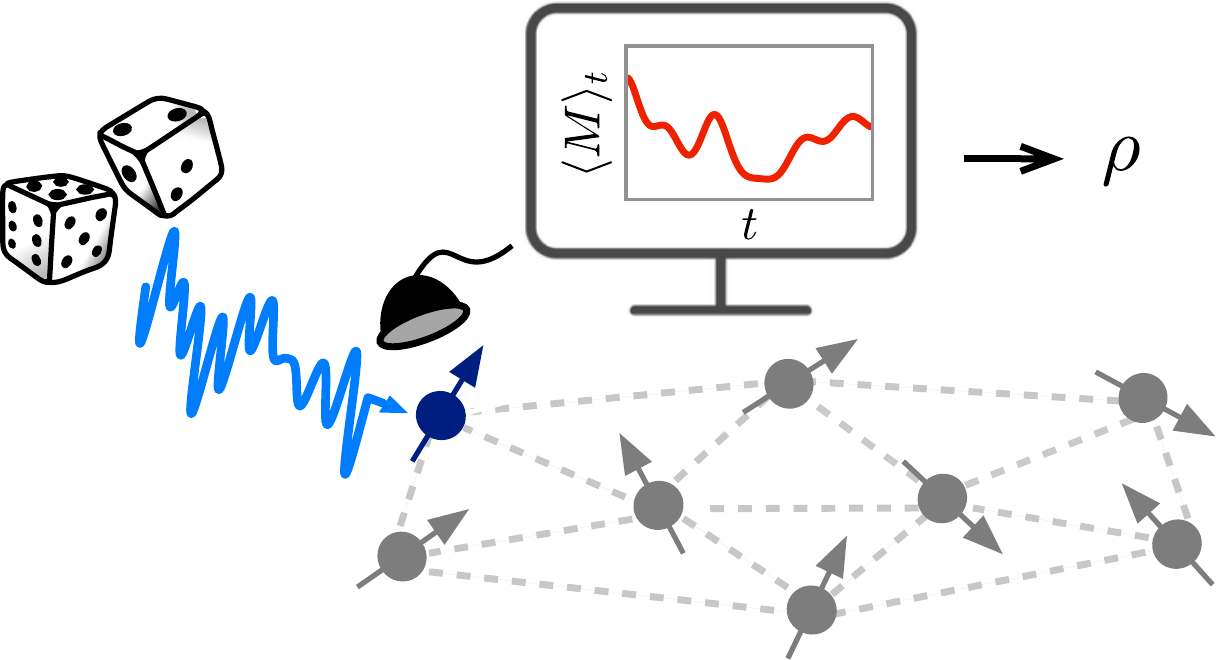}
	\caption{\label{fig1}Random-field quantum-state tomography of a multiqubit system when only a single qubit (dark blue) can be accessed. By randomly driving (light blue) the qubit and measuring the expectation of any single-qubit observable $M$, for a fully controllable system and sufficiently long times of the signal $\langle M\rangle_{t}$ (red), any qubit-network state $\rho$ can almost always be reconstructed. We experimentally demonstrate this principle by reconstructing combined states of an electron-nuclear spin system in diamond, in which only the electron spin is accessible.} 
\end{figure} 
  
Here, we provide a solution to the drawbacks of (i) and (ii) through the observation that a \emph{random} control field can create a random unitary evolution~\cite{banchi2017driven} when the system is \emph{fully controllable}, i.e., when there exist pulse shapes that, in principle, allow every unitary evolution to be created~\cite{d2007introduction}. We show that in this case, a randomly applied field (almost always) yields enough information in the measurement signal of any observable to reconstruct an arbitrary quantum state, provided the   signal is long enough. Thus, for qubit systems that are fully controllable by addressing a single qubit, a local random pulse, that randomly "shakes" the total system, allows for the reconstruction of the full state of the qubit network by measuring only a \textit{single-qubit} observable (see Fig.~\ref{fig1}). We experimentally demonstrate this principle in a solid-state spin system in diamond, but due to its generality, the presented random-field-based tomography  constitutes a broadly applicable strategy that can be readily adopted in a variety of partially-accessible systems.\\ 

\textit{Theory.--} Adopting the framework of~\cite{silberfarb2005quantum,merkel2010random}, we begin by developing the theory behind random-field quantum-state tomography. While the following assessment is completely general, for the sake of simplicity, we restrict ourselves to a single random field.  

Consider a $d$-dimensional quantum system initially in an unknown state $\rho$, whose evolution is governed by a time-dependent Hamiltonian of the form 
\begin{align}
\label{eq:Ham}
H(t)=H_{0}+f(t)H_{c},
\end{align}
 that depends on a classical control field $f(t)$ steering the system. The time evolution of the expectation of an observable $M$ is then given by
\begin{equation}
\label{eq:evoobservable}
 \langle M\rangle_{t}=\text{Tr}\{U^{\dagger}_{t}MU_{t}\rho\},
  \end{equation}
 where $U_{t}=\mathcal T\exp\{-i\int_{0}^{t}H(t^{\prime})dt^{\prime}\}$ is the time-evolution operator in units of $\hbar= 1$, with $\mathcal{T}$ indicating time ordering. We assume, without loss of generality, that $M$ is traceless. The quantum system is said to be fully controllable if there exist pulse shapes that allow for creating every unitary evolution. For unconstrained control fields this is guaranteed if and only if the dynamical Lie algebra $\mathfrak L=\text{Lie}(iH_{0},iH_{c})$ generated by nested commutators and real linear combinations of $H_{0}$ and $H_{c}$ spans the full space (i.e., $\mathfrak{u}(d)$ or $\mathfrak{su}(d)$ for traceless Hamiltonians)~\cite{d2007introduction}. Using the generalized Bloch-vector representation, we can write the initial state as $\rho=\frac{\mathds{1}}{d}+\sum_{m=1}^{d^{2}-1}r_{m}B_{m}$, where $\mathds{1}$ denotes the identity and $\mathbf{r}=({\rm Tr}\{\rho B_1\},\cdots,{\rm Tr}\{\rho B_{d^{2}-1}\})$ is the Bloch vector, with $\{B_{m}\}_{m=1}^{d^{2}-1}$ being a complete and orthonormal basis for traceless and Hermitian operators. This allows for \eqref{eq:evoobservable} to be expressed as $\langle M\rangle_{t}=\sum_{m=1}^{d^{2}-1} A_{t,m}r_{m}$, where $A_{t,m}=\text{Tr}\{U_{t}^{\dagger}MU_{t}B_{m}\}$. We assume that at times $t=n\Delta t$, with $n=1,\cdots, (d^{2}-1)$, the expectation $\langle M\rangle_{t}$ is measured, so that we obtain $d^{2}-1$ values, which are collected in the vector $\mathbf{y}\equiv(\langle M\rangle_{\Delta t},\cdots,\langle M\rangle_{(d^{2}-1)\Delta t})$, referred to as the measurement record. The measurement record is determined by the set of equations 
 \begin{align}
 \label{eq:record}
 \mathbf{y}=\mathcal M[f(t)]\mathbf{r},
 \end{align}
 where we have indicated, here, the explicit dependence of the matrix $\mathcal M\in \mathbb R^{(d^{2}-1)\times (d^{2}-1)}$, with entries given by $\mathcal M_{n,m}=A_{n\Delta t,m}$, on the control field $f(t)$. We call the measurement record \emph{informationally complete} if $\mathcal M$ is invertible, thereby allowing the state $\rho$ to be inferred via $\mathbf{r}=\mathcal M^{-1}\mathbf{y}$.
  
How can we ensure that the field and the measurement intervals chosen allow for inverting $\mathcal M$? It can be seen that if the system is not fully controllable, which is equivalent to the existence of symmetries~\cite{zimboras2015symmetry}, not every $\rho$ can be reconstructed~\cite{merkel2010random}. In contrast, for fully controllable quantum systems it is, in principle, possible to determine the pulses that create an informationally complete measurement record. For instance, this can be achieved through optimal-control algorithms designed to identify control fields that rotate $M$ into $\{B_{m}\}$, so that $\mathcal M$ is diagonal. However, optimal control typically depends on the availability of an accurate model. Moreover, it can be computationally expensive, and the designed pulses are often challenging to implement in the laboratory. 

Fortunately, it was recently shown that for fully controllable systems a Haar-random unitary evolution (i.e., unitary transformations that are uniformly distributed over the unitary group~\cite{banchi2017driven}) is created when $f(t)$ is applied at random over an interval $[0,T_{*}]$ \cite{banchi2017driven}. The Haar-random time $T_{*}$ can be estimated from the time required to converge to a unitary $t$-design, which can be accomplished by mapping the expected evolution to the dynamics generated by a Lindbladian and finding its gap \cite{banchi2017driven}. Thus, a random field of length $(d^{2}-1)T_{*}$ along with measurements of the expectation of $M$ at time intervals $\Delta t=T_{*}$, yields row vectors of $\mathcal M$ that are statistically independent, due to the unitary invariance of the Haar measure. Furthermore, since the row vectors are uniformly distributed, with unit probability they are also linearly independent. This leads to the result that for almost all random pulse shapes, but a set of measure zero, the matrix $\mathcal M$ is invertible. Hence, almost all pulse shapes allow for reconstructing $\rho$ by measuring the expectation of any observable $M$. With further details found in the Supplemental Material~\cite{supp}, we  summarize these findings in the following theorem. \\

\textit{Theorem.} For a $d$-dimensional fully controllable quantum system subject to a random field of length $t=(d^{2}-1)T_{*}$, with $T_{*}$ being the Haar-random time, the measurement record of any observable $M$ determined by \eqref{eq:record} with $\Delta t=T_{*}$ is almost always informationally complete. \\ 

\begin{figure*}
\centering
	\includegraphics[width=0.98\linewidth]{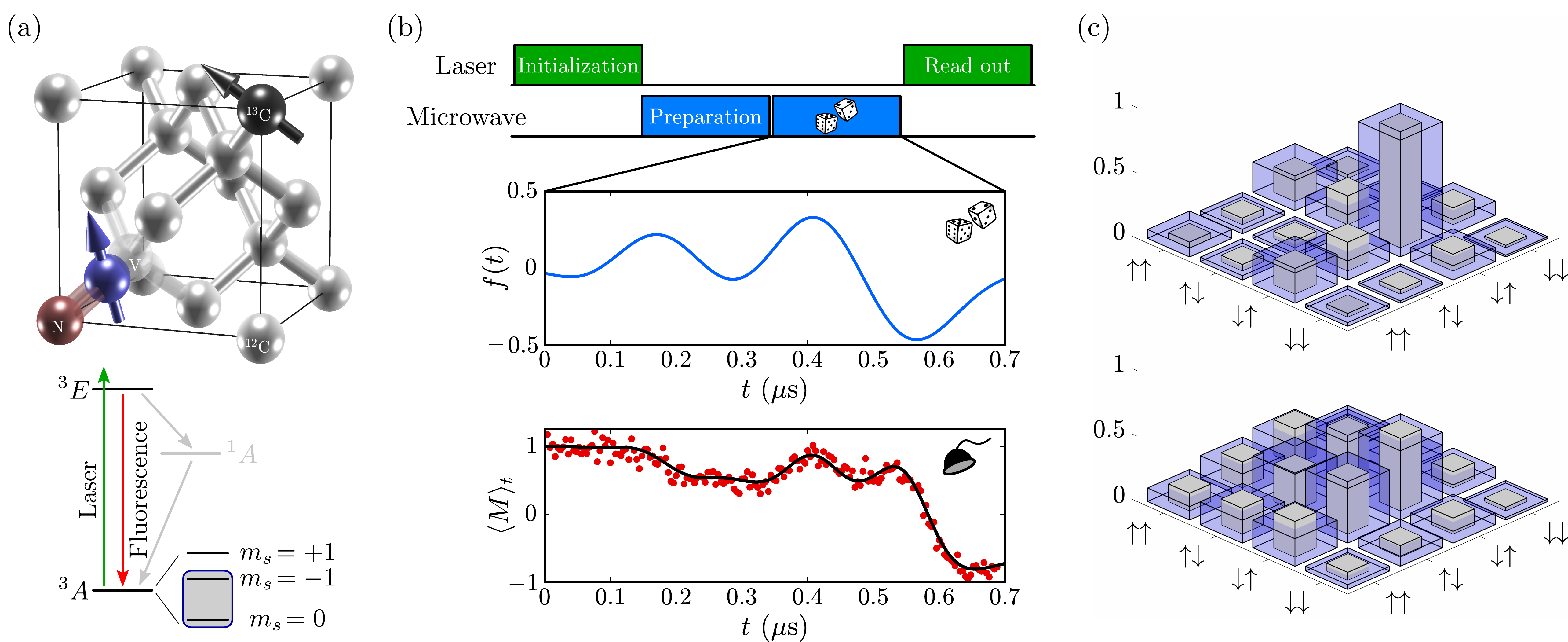}
	\caption{\label{fig2}(a)  Schematic illustration of the two-spin system in diamond: NV-center electron spin (dark blue) and nearby carbon nuclear spin (black), along with the level diagram of the electron spin including the fine structure of the ground state, shown below. (b) Sequence of the applied fields: laser for initialization and readout (green), microwave for state preparation and tomography (blue). The central panel shows an example random pulse shape $f(t)$ and the corresponding time trace $\langle M\rangle_{t}$ of the electronic ground-state population (bottom panel), when the preparation stage is absent, namely the system is initially only in the polarized state. The solid black line shows the ideal case obtained by numerical propagation of the initial state. (c) Random-field reconstruction of two different entangled states: both figures show the modulus of the reconstructed density matrix, wherein the solid gray bars depict the tomography results and the blue transparent bars show the exact state obtained numerically. The upper panel shows the reconstruction with fidelity $96.1\%$ of a randomly created state, whereas the lower panel shows the reconstruction with fidelity $94.9\%$ of a highly entangled state (concurrence $C=0.91$) created through an optimized preparation pulse shape.}  
\end{figure*}

Since full controllability can often be obtained by acting with a single control $H_{c}$ on a part of the system only, e.g., a single qubit \cite{heule2010local, burgarth2010scalable, arenz2014control, zeier2011symmetry, schirmer2008global}, the appeal of this theorem is twofold: under the premise of full controllability, arbitrary quantum states can almost always be reconstructed (i) without the need for expensive numerical pulse designs and (ii) requiring only partial system access. Furthermore, full controllability of systems of the form~\eqref{eq:Ham} is a generic property, as almost all $H_{0}$ and $H_{c}$ generate the dynamical Lie algebra $\mathfrak{L}$~\cite{altafini2002controllability,wang2016subspace}. This leads to the general corollary: \\

\textit{Corollary.} Full quantum-state tomography of \textit{almost all} randomly-driven quantum systems of the form~\eqref{eq:Ham} is possible by reading out a single observable.  \\  

We remark that the above should be treated as a mathematical fact rather than a source of physical intuition. Nevertheless, it should be noted that in cases where full control is not achieved with a single field, adding additional control fields can be a straightforward approach for obtaining full controllability. In fact, if full system access is possible, in fully connected qubit networks two controls on each qubit are sufficient~\cite{zeier2011symmetry}. In general, a variety of algebraic tools and criteria \cite{d2007introduction,burgarth2009local}, as well as numerical algorithms \cite{zimboras2015symmetry, schirmer2001complete}, can be used to determine whether full control is achieved with the control field(s) at hand.  Even in situations where full control is not achievable, as long as the state and the observable lie within the span of the dynamical Lie algebra, we expect random-field quantum-state tomography to succeed.\\ 

 \textit{Experiment.--} In order to demonstrate the utility of the above principle, we experimentally perform the random-field tomography of a system of two interacting qubits ($d=4$). The solid-state spin system we employ is depicted in Fig.~\ref{fig2}(a) and consists of the electron spin of a nitrogen-vacancy (NV) center in diamond~\cite{doherty13}, coupled to the nuclear spin of a nearby $^{13}$C atom via hyperfine interaction. In the ground-state triplet, the NV center has the electronic sublevels $m_s=0,\pm1$, where the degeneracy between the $m_s=\pm1$ states is lifted by a  magnetic field  of strength $B\approx 504.7$~G along the NV axis. The first qubit is formed by the $m_s=0$ state, denoted by $\ket{\uparrow}_1$, and the $m_s=-1$ state, denoted by $\ket{\downarrow}_1$, of the electron spin [see lower panel of Fig.~\ref{fig2}(a)]. Likewise, for the second qubit we denote the $^{13}$C nuclear spin states with quantum numbers $m_I=\pm 1/2$ by $\ket{\uparrow}_2$ and $\ket{\downarrow}_2$, respectively. Furthermore, we represent the Pauli operators of the two qubits by $\sigma_j^\kappa$, for $j=1,2$ and $\kappa=x,y,z$, where $\ket{\uparrow}_j$ and $\ket{\downarrow}_j$ are the $\pm 1$ eigenstates of $\sigma_j^z$, respectively. In a rotating frame such a system is described by the Hamiltonian~\cite{supp}
\begin{equation}
\label{eq:drift}
H_0=\frac{\omega_1}{2}\sigma_1^z+\frac{\omega_2}{2}\sigma_2^z+\frac{\Omega_2}{2}\sigma_2^x+\frac{g_z}{2}\sigma_1^z\sigma_2^z+\frac{g_x}{2}\sigma_1^z\sigma_2^x.
\end{equation}
Since the gyromagnetic ratio of the nuclear spin is three orders of magnitude smaller than that of the electron spin, access to the system is effectively restricted to the electron spin, as a direct read out of the nuclear spin is extremely challenging. The electron spin is driven through a classical field, whose coupling to the electron spin is described by the control Hamiltonian
\begin{align}
\label{eq:control}
  H_c=\frac{\Omega_1}{2} \sigma_1^x.
\end{align}
This control is achieved by applying a microwave field of frequency $\omega$, which is generated by an arbitrary waveform generator (AWG) and delivered to the sample through a copper microwave antenna, after being amplified by a microwave amplifier. The precise control over the AWG allows us to engineer the control field $f(t)$ with arbitrary amplitude modulations. The control field amplitude $\Omega_1$ is calibrated with the output power of the AWG by measuring the frequency of Rabi oscillations of the electron spin~\cite{supp}, i.e., for $f(t)\equiv 1$. We choose a microwave frequency $\omega/2\pi\approx 1455.5$~MHz, which, under the applied magnetic field, lies between the two allowed transitions between eigenstates of $H_0$~\cite{dreau2012}. The parameters in our experiment are $\{\omega_1,\omega_2,\Omega_1,\Omega_2,g_z,g_x\}/2\pi=\{-2.97,-6.46,7.91,-1.39,5.92,1.39\}$~MHz,
with minor variations~\cite{supp}, e.g., due to small drifts in the magnetic field between different runs of the experiment, which leads to imperfections in the state preparation. 

A system described by the Hamiltonian~\eqref{eq:Ham}, with $H_{0}$ and $H_{c}$ defined in~\eqref{eq:drift} and \eqref{eq:control}, respectively, is fully controllable, as the dynamical Lie algebra spans $\mathfrak{su}(4)$. As an observable we choose the population of the electronic $m_s=0$ state, represented by $M=\sigma_1^z$, which can easily be read out by state-dependent fluorescence~\cite{doherty13}. To create a random control field we design random pulse shapes $f(t)$ based on a truncated Fourier series~\cite{banchi2017driven}
\begin{equation}
\label{eq:random_pulse}
f(t)=\sum_{j=1}^{K}F_j\cos(\nu_jt+\varphi_j),
\end{equation}
with uniformly-distributed random variables: amplitudes $F_j$ (fulfilling the normalization $\sum_{j=1}^KF_j=1$), frequencies $\nu_j/2\pi\in[0,4]$~MHz, and phases $\varphi_j\in[0,2\pi]$. Because of a limited coherence time, instead of using a single random pulse shape, in the experiment we use $d^{2}-1=15$ separate random pulses to create linearly independent rows of $\mathcal M$, thereby only evolving the system up to a time $\Delta t$ in each run. Throughout the remainder, we employ random pulses with $K=10$ Fourier components and a length of $\Delta t=0.7~\mu$s (see Supplemental Material Fig.~S1), which lies well below the coherence time of the microwave-driven system, and also allows for moderate levels of noise in the measurement record~\cite{supp}.

As a first check of the random-field tomography we reconstruct the state after the optical ground-state polarization with a 532~nm laser, i.e., with an empty preparation stage in Fig.~\ref{fig2}(b), which ideally leads to the pure state $\rho_{0}=\ket{\psi_0}\bra{\psi_0}$, with $\ket{\psi_0}=\ket{\uparrow}_1\ket{\uparrow}_2$~\cite{fischer2013}. To reconstruct this state we consecutively apply $15$ random pulses on the electron spin. One example random pulse shape $f(t)$ is shown in the central panel of Fig.~\ref{fig2}(b), with the corresponding full time trace of the expectation value $\langle M\rangle_t$ depicted below. In order to reconstruct the density operator from the obtained measurement record, we employ a least-square type minimization~\cite{supp}, using the last 10 data points of every random pulse [see Supplemental Material Fig.~S2(a)]. The resulting reconstructed density matrix yields $97.7\%$ fidelity with $\rho_0$ [see Supplemental Material Fig.~S2(b)]. 

In order to demonstrate the reconstruction of nontrivial states, as a first example, we randomly create a state by applying a preparation pulse of the form~\eqref{eq:random_pulse} with a duration of 0.8~$\mu$s [see Supplemental Material Fig.~S3(a)], after the initialization of the system into the state $\rho_{0}$. Since we have to preform 15 tomography pulses, and the slight drift in the experimental parameters leads to small differences in the states created from $\rho_0$ through the preparation stage before each of these pulses, the resulting state shows some impurity. The modulus of the reconstructed density matrix is shown in the upper panel of Fig.~\ref{fig2}(c) (gray bars). The reconstructed state shows a $96.1\%$ fidelity with the state ideally prepared (blue transparent bars) under the random pulse. The entanglement of this state, as quantified by the concurrence $C$, is given by $C=0.48$. As another example, we optimize a preparation pulse of the form~\eqref{eq:random_pulse} with a pulse length $1.8~\mu$s [see Supplemental Material Fig.~S3(b)] to create a highly entangled state of the two-qubit system. The modulus of the obtained density matrix, which also shows some impurity due to the preparation before each tomography pulse, is depicted in the lower panel of Fig.~\ref{fig2}(c). The reconstructed state has a concurrence $C=0.91$ and shows a fidelity of $94.9\%$ with the ideally prepared state. In the latter two cases the ideal states are obtained by numerical propagation of the initial state $\rho_0$ under the preparation pulses. \\

\textit{Discussion.--} We have shown that by randomly driving and measuring a single component of a multipartite quantum system, the quantum state of the total system can be reconstructed. This is a consequence of the fact that the data collected through expectation measurements of a single observable almost always contain enough information to reconstruct any state, provided the system is fully controllable and the randomly applied field is long enough. Based on this principle, we presented the successful experimental creation and reconstruction of composite states of an NV-center electron spin and a nuclear spin in diamond with high fidelities. The exponential overhead needed to reconstruct generic quantum states of qubit systems is reflected in the $d^{2}-1$ expectation measurements, as well as in in the length of the random pulse. However, numerical evidence presented in Fig.~S1 of the Supplemental Material suggests that often pulses much shorter than $(d^2-1)T_{*}$ can yield information completeness. Further, we  remark that, for low-rank quantum states, we expect that the number of expectation measurements required can also be significantly reduced when random-field tomography is combined with compressed sensing methods~\cite{gross2010quantum, flammia2012quantum,ohliger2013efficient, kalev2015quantum, shabani2011efficient, christandl2012reliable,riofrio2017experimental, steffens2017experimentally, cramer2010efficient, kalev2015quantum}. It is also worth mentioning that in other settings, the knowledge of the full quantum state may not be necessary; instead, information carried in expectations of  only certain many-body operators may be desired~\cite{guehne2002}. For example, this is the case in hybrid quantum simulation~\cite{mcclean2016theory}, where such expectation measurements are used by a classical co-processor to update a set of parameters governing the quantum simulation~\cite{kokail2019self, hempel2018quantum, peruzzo2014variational, kandala2017hardware,elben2019zoller}. We believe that a variant of the presented random-field approach could offer a way to extract the desired information with reduced overhead in accessing the system.

Besides full controllability, we also assumed knowledge of the model describing the controlled system. This assumption was needed to \emph{numerically} calculate the unitary evolution $U_{t}$ in~\eqref{eq:record}, which allowed for calculating $\mathcal M$. However, this assumption is not crucial, given that \emph{process tomography} can be performed without any prior knowledge of the model~\cite{burgarth2012quantum,blume2013robust}. That is, instead of numerically calculating $U_{t}$, the unitary evolution can \emph{experimentally} be determined. This can be achieved by additionally creating a complete set of states, for instance through randomly rotating the unknown state $\rho$. Since under the premise of full controllability uniformly-distributed states can be created through a random pulse shape, this implies that state and process tomography are possible by randomly driving and measuring a single system component without knowing system details. Therefore, the price is an increase in the number of expectation measurements needed,  estimated to be $\mathcal O(d^{4})$~\cite{blume2013robust,hou2019}. However, the observation that no prior knowledge except full controllability is needed raises an interesting prospective: it is possible to fully control and read out a quantum system only based on measurement data~\cite{judson1992teaching,chen2018combining,li2017hybrid} by accessing merely part of the system~\cite{lloyd2004universal}.  As such, under the premise of full controllability, a quantum computer/simulator can, in principle, be fully operated by processing classical data obtained from randomly driving and measuring a single qubit without knowing the physical hardware the quantum computer/simulator is made off.\\

\noindent
\textit{Acknowledgements.--} The authors thank R. Kosut, A. Magann, B. G. Taketani, and J. M. Torres for helpful comments. This work is supported by the National Natural Science Foundation of China (Grants No.~11950410494, No.~11574103, No.~11874024), the National Key R$\&$D Program of China (Grant No.~2018YFA0306600), and the Fundamental Research Funds for the Central Universities. Furthermore, C.A. is supported by the ARO (Grant No.~W911NF-19-1-0382).\\



%

\clearpage

\onecolumngrid
\begin{center}
	\textbf{\large Supplemental Material: \\
		Complete Quantum-State Tomography with a Local Random Field}
\end{center}	
\vspace{5ex}
\twocolumngrid

\section{Proof of the theorem}
Here, we show that if at the measurement times $t=nT_{*}$, with $n=1,\cdots,(d^{2}-1)$, statistically independent Haar-random unitaries $U_{n}$ are created, then with probability 1 the matrix $\mathcal M$ with entries $\mathcal M_{n,m}=\text{Tr}\{U_{n}^{\dagger}MU_{n}B_{m}\}$ is invertible. This result should not be too surprising, since it is well known in the literature that observables created through Haar-random unitary operations are almost always informationally complete (see, e.g., \cite{ohliger2013efficient,merkel2010random}).   

 We first note that if $U_{n}$ is Haar random, then the Hermitian matrices $U_{n}^{\dagger}M U_{n}$ are independent and uniformly distributed within the set of Hermitian matrices, with the same spectrum as $M$. Equivalently, the row vectors $a_{n}=(\mathcal M_{n,1},\cdots, \mathcal M_{n,d^{2}-1})$ of $\mathcal M$ are independent and uniformly distributed within a vector space $V_{\text{Her}}$. For an explicit characterization of the measure induced on $V_{\text{Her}}$ we refer to~\cite{ohliger2013efficient}. We proceed by recalling a standard result from measure theory: the probability that choosing $d^{2}-1$ independent vectors $a_{n}$ uniformly random (on $V_{\text{Her}}$) are linearly independent is 1, which implies that $\mathcal M$ is invertible with probability 1. For completeness we give the proof below. Clearly, the first vector $a_{1}$ is linearly independent with probability 1. Now assume that the vectors $a_{1},\cdots, a_{d^{2}-2}$ are linearly independent with probability 1, i.e., assume that they span a $d^{2}-2$ dimensional subspace $V$, which intersects $V_{\text{Her}}$. Note that $V$ has measure zero within $V_{\text{Her}}$. Thus, the probability that $a_{d^{2}-1}\notin V$ is 1. Consequently, the probability that $\text{span}\{a_{1},\cdots,a_{d^2-1}\}=V_{\text{Her}}$ is 1, which completes the proof.

\section{System Hamiltonian}
In the NV-center ground-state manifold [the state with $^3A$ symmetry, see Fig.~2(a)] the Hamiltonian can be written as~\cite{doherty13,dreau2012}
\begin{equation}
\label{eq:Hgs}
H_\text{gs}=DS_z^2-\gamma_eBS_z+H_{\text N} -\gamma_{\text C}BI_z+\mathbf{S}\cdot\boldsymbol{\mathcal{A}}\cdot\mathbf{I},
\end{equation}
with the zero-field splitting $D/2\pi=2.87$~GHz. The electron spin-1 and $^{13}$C nuclear spin-1/2 operators are denoted by $\mathbf{S}$ and $\mathbf{I}$, and their respective gyromagnetic ratios are given by $\gamma_e/2\pi=-2.8~$MHz/G and $\gamma_{\text C}/2\pi=1.07~$kHz/G. The Hamiltonian $H_{\text N}$ describes the intrinsic nitrogen nuclear spin of the NV center, including its hyperfine interaction with the electron spin. The nitrogen nuclear spin is polarized into its $m_I=+1$ state during the optical ground-state polarization of the electron spin using a 532~nm green laser pulse~\cite{fischer2013} and remains therein afterward. The part $H_{\text{N}}$ thereby only entails an energy shift $\mathcal{A}_\text{N}/2\pi=\pm 2.16$~MHz of the electronic $m_s=\pm1$ states  due to the hyperfine interaction~\cite{felton2009,dreau2012}. The hyperfine tensor $\boldsymbol{\mathcal{A}}$, on the other hand, describes the dipole-dipole interaction between the NV-center electron spin and the carbon nuclear spin. In a secular approximation and with an appropriate $x$ axis we can write this interaction as $\mathbf{S}\cdot\boldsymbol{\mathcal{A}}\cdot\mathbf{I}=\mathcal{A}_{zz}S_zI_z+\mathcal{A}_{zx}S_zI_x$.

Furthermore, the coupling of the electron spin to the microwave field can be described by the Hamiltonian $H_{\text{mw}}(t)=\sqrt{2}f(t)\Omega_1\cos(\omega t)S_x$ in the lab frame, where the factor $\sqrt{2}$ is included for convenience. In the subspace spanned by the $m_s=0$ and $m_s=-1$ states, for the electron-spin operators one can make the two substitutions $S_x\rightarrow\sigma_1^x/\sqrt{2}$ and $S_z\rightarrow(\sigma_1^z-\mathds{1}_1)/2$. Going to an interaction picture with respect to $\omega\sigma_1^z/2$ and performing a rotating-wave approximation, the system Hamiltonian $H_{\text{gs}}+H_{\text{mw}}(t)$ then takes the form~(4), with $\omega_1=\omega-D+\gamma_e B+\mathcal{A}_{\text N}$, $\omega_2=\gamma_nB-\mathcal{A}_{zz}/2$, $\Omega_2=-\mathcal{A}_{zx}/2$, $g_z=\mathcal{A}_{zz}/2$, and $g_x=\mathcal{A}_{zx}/2$. In our sample, we find a longitudinal hyperfine coupling $\mathcal{A}_{zz}/2\pi=(11.832\pm0.005)$~MHz and a transversal coupling $\mathcal{A}_{zx}/2\pi=(2.790\pm0.002)$~MHz. This was shown in electron-spin resonance (ESR) experiments~\cite{yu2018pub} and yields the parameters given in the main text. Drifts in the magnetic field and microwave amplitude between different runs of the experiment lead to
$\{\omega_1,\Omega_1\}/2\pi\in\{-2.974\pm0.247,7.910\pm0.129\}$~MHz, where in every separate run they can be determined up to a precision of a few kHz.

\section{Tomography using noisy data} 
\begin{figure}[h]
	\includegraphics[width=0.95\linewidth]{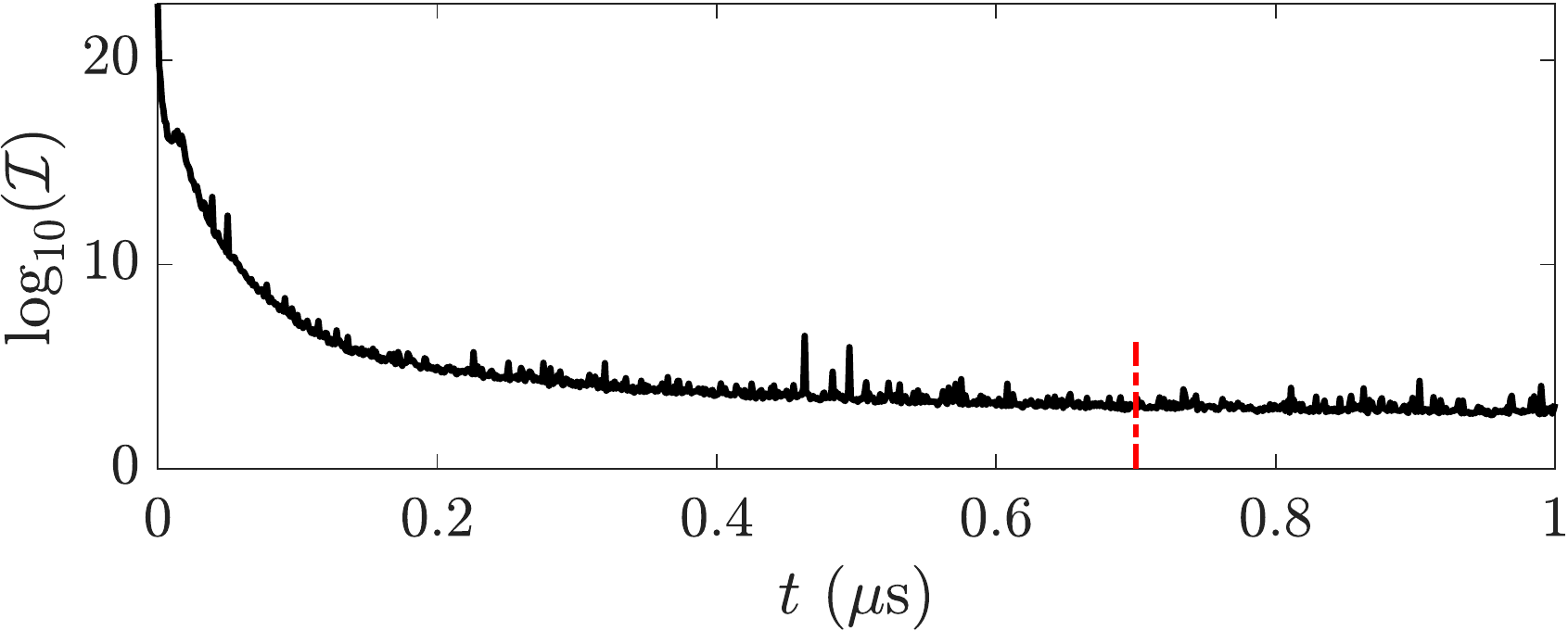}
	\caption{\label{fig3}Time evolution of the logarithm of the average over $||\mathcal{M}^{-1}||$ from 1000 random pulses using the experimental parameters. The red dashed line shows the random-pulse length $0.7~\mu$s we employ in our tomography.}   
\end{figure}

While in theory the measurement record $\mathbf{y}$ allows a perfect reconstruction of $\rho$, in practice the expectation measurements always contain  noise described by some error $\boldsymbol{\epsilon}$, that adds to $\mathbf{y}$. A common way to reconstruct $\rho$ from noisy data $\tilde{\mathbf{y}}=\mathbf{y}+\boldsymbol{\epsilon}$ is to find the Bloch vector $\mathbf{r}$ that best explains the measured data $\tilde{\mathbf{y}}$. In order to do so, we solve the optimization problem
\begin{equation}
\label{eq:minimization}
\arg \min_{\mathbf{r}}\Vert\tilde{\mathbf{y}} -\mathcal M[f(t)]\mathbf{r} \Vert,   
\end{equation}
such that $\rho\geq 0$ and $\text{Tr}\{\rho\}=1$. This optimization is done simultaneously for multiple measurement outcomes $\tilde{\mathbf{y}}^{(j)}$ and corresponding matrices $\mathcal{M}^{(j)}$, e.g., with $j=1,...,10$ in our reconstructions, as shown in Fig.~\ref{fig3}(a). The explicit operator basis $\{B_m\}_{m=1}^{d^2-1}$ that we employ for the representation of $\mathcal{M}$ and $\mathbf{r}$ in our state reconstructions is the standard Pauli-operator basis.

Note that since the error between the Bloch vector $\mathbf{r}$ corresponding to the noiseless measurement record and the one obtained from noisy data is given by $\Vert \mathcal M^{-1}\boldsymbol{\epsilon}\Vert$, pulse shapes minimizing $\Vert \mathcal M^{-1}\Vert$ allow for more robust state reconstruction, while pulses yielding a matrix $\mathcal M$ close to singular are more sensitive to measurement errors. More sophisticated statistical methods as well as different objective functions optimized over the control fields can be employed to design control fields that achieve the best performance for noise-robust quantum-state tomography. We remark here that it is well known that state reconstruction through randomly-created observables is already surprisingly robust against moderate levels of noise~\cite{candes2011probabilistic}. We study this robustness by analyzing $||\mathcal{M}^{-1}||$ for the experimental setting at hand. Therefore, in Fig.~\ref{fig3} we show the time evolution of the quantity $\mathcal{I}$, by which we denote the average of $||\mathcal{M}^{-1}||$ over 1000 realizations of the random pulses~(6), calculated numerically using the parameters of the experiment. The norm of the inverse saturates and the dashed red line indicates the pulse length we employ in our experimental state tomography, since we see no significant decrease in $\mathcal{I}$ after $0.7~\mu$s.

\section{Initial-state reconstruction}
\begin{figure}[t]
	\includegraphics[width=0.95\linewidth]{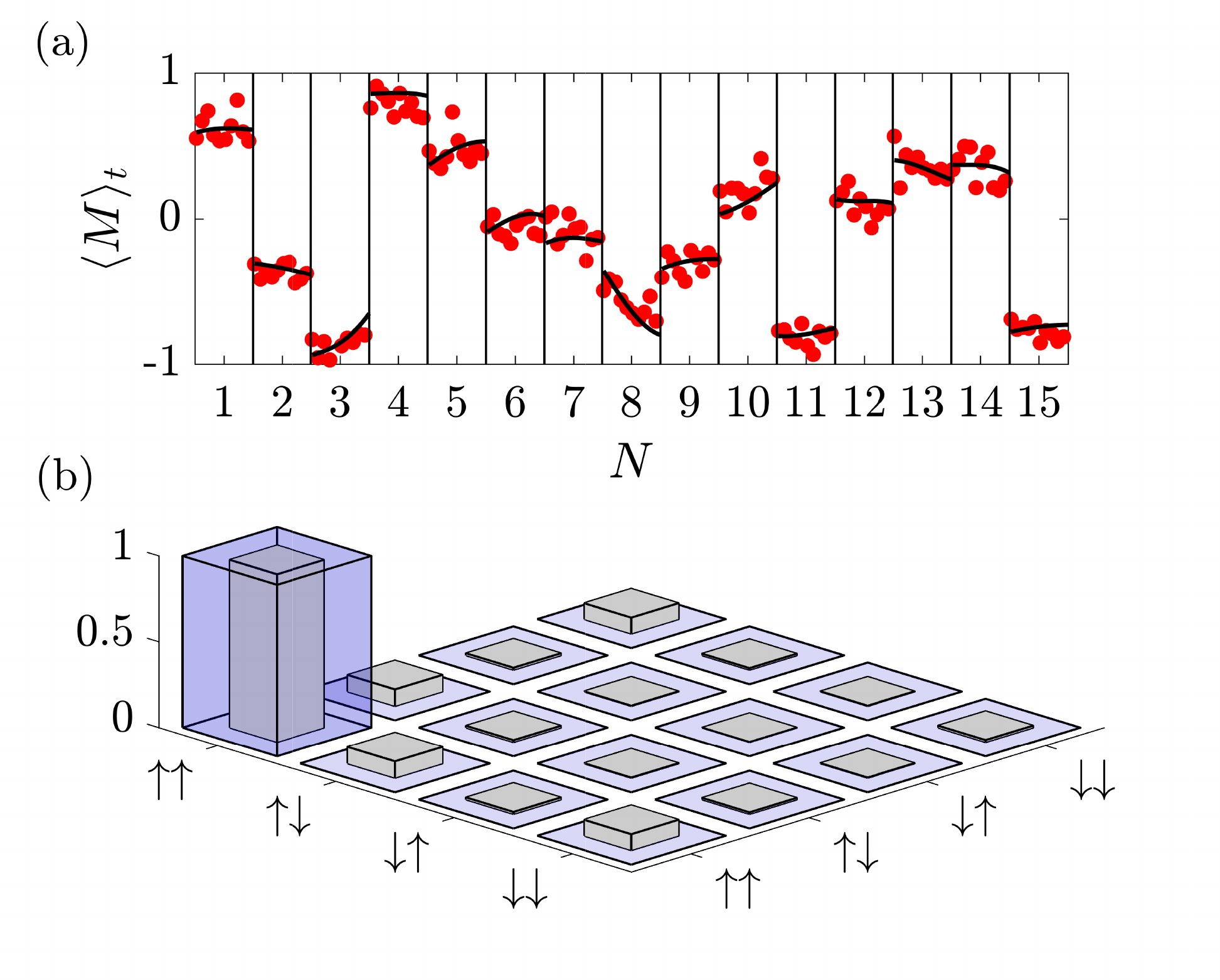}
	\caption{\label{fig4}(a) Last 10 data points of the expectation measurement obtained under the $15$ random pulses (labeled by $N$). (b) Modulus of the reconstructed density matrix (gray bars) and of the exact state $\rho_0$=$\ket{\psi_0}\bra{\psi_0}$ (blue bars), with $\ket{\psi_0}=\ket{\uparrow}_1\ket{\uparrow}_2$, yielding  a $97.7\%$ reconstruction fidelity.}   
\end{figure}

 For the reconstruction of the density matrices we use the vectors $\tilde{\mathbf{y}}^{(j)}$ obtained from the last 10 data points of the expectation measurements under the 15 random pulses, along with the corresponding matrices $\mathcal{M}^{(j)}$ ($j=1,...,10$) and perform the minimization~\eqref{eq:minimization}. As mentioned in the main text, as a first check we perform the tomography of the state $\rho_0$ after the optical ground-state polarization, i.e., when the preparation stage in Fig.~2(b) is void. The corresponding measurement data and numerical propagation results are shown in Fig.~\ref{fig4}(a). The reconstructed density matrix is depicted as gray bars in Fig.~\ref{fig4}(b) and yields a $97.7\%$ fidelity with the ideal state $\rho_0$, which is indicated by blue transparent bars for comparison.

\section{Preparation pulses} 
\begin{figure}[h]
	\includegraphics[width=0.95\linewidth]{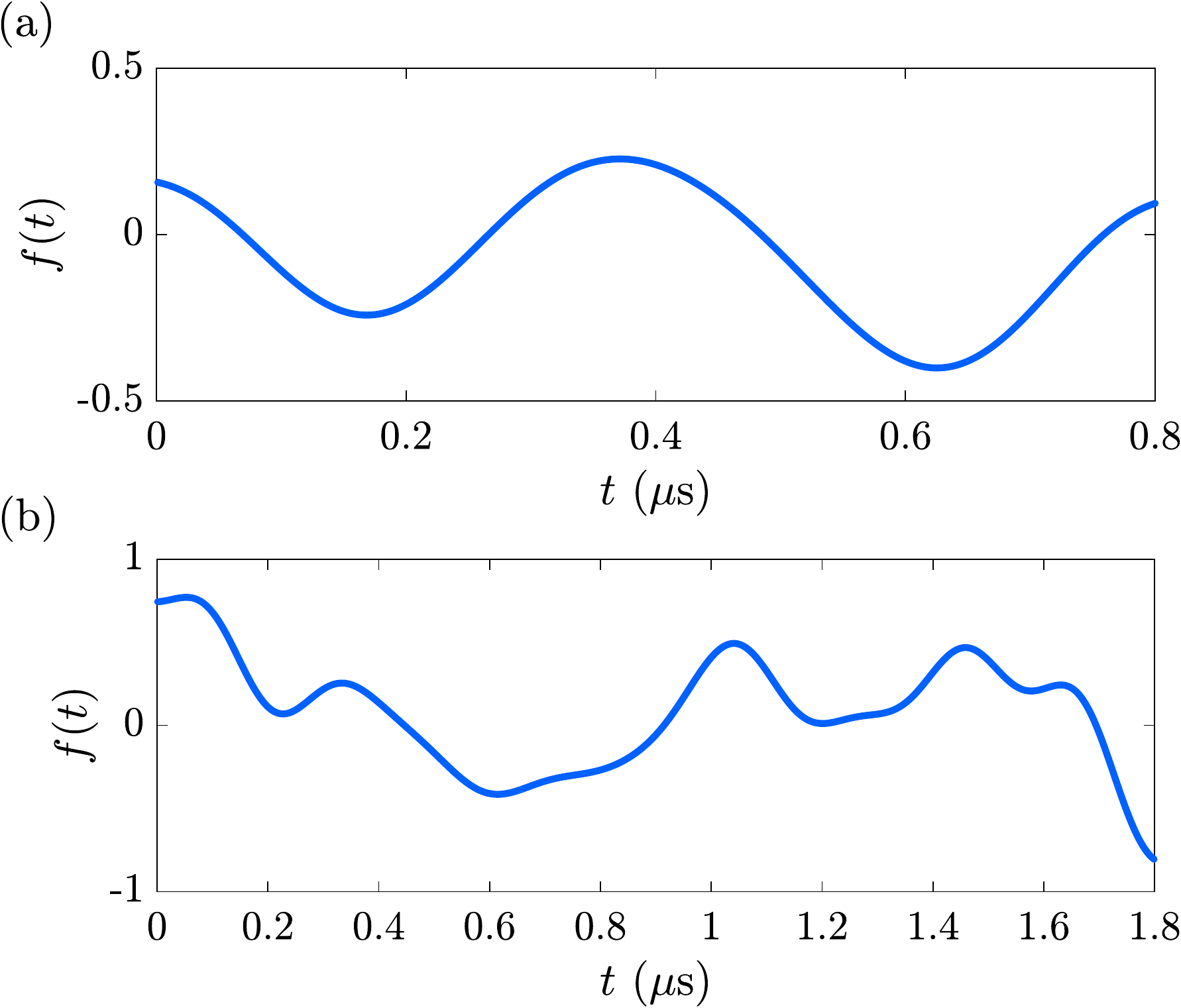}
	\caption{\label{fig5}(a) Pulse shape for the preparation of the random state shown in the upper panel of Fig.~2(c). (b) Numerically optimized preparation pulse shape to create the highly entangled state (concurrence $C=0.91$) shown in the lower panel of Fig.~2(c).}  
\end{figure}

The two non-trivial states we reconstructed in Fig.~2(c) are prepared using the microwave preparation stage  [see top panel of Fig.~2(b)]. For the sake of completeness, the pulse shapes of the two preparation pulses are depicted in Fig.~\ref{fig5}. Here, Fig.~\ref{fig5}(a) shows the random pulse employed to create the state shown in the upper panel of Fig.~2(c). The pulse shape for the creation of the highly entangled state shown in the lower panel of Fig.~2(c) was obtained by numerical optimization of the truncated Fourier series~(6), in order to achieve the highest concurrence for a fixed preparation time of 1.8~$\mu$s, and is depicted in Fig.~\ref{fig5}(b).

\section{Experimental parameters} 
\begin{figure}[t]
	\includegraphics[width=0.95\linewidth]{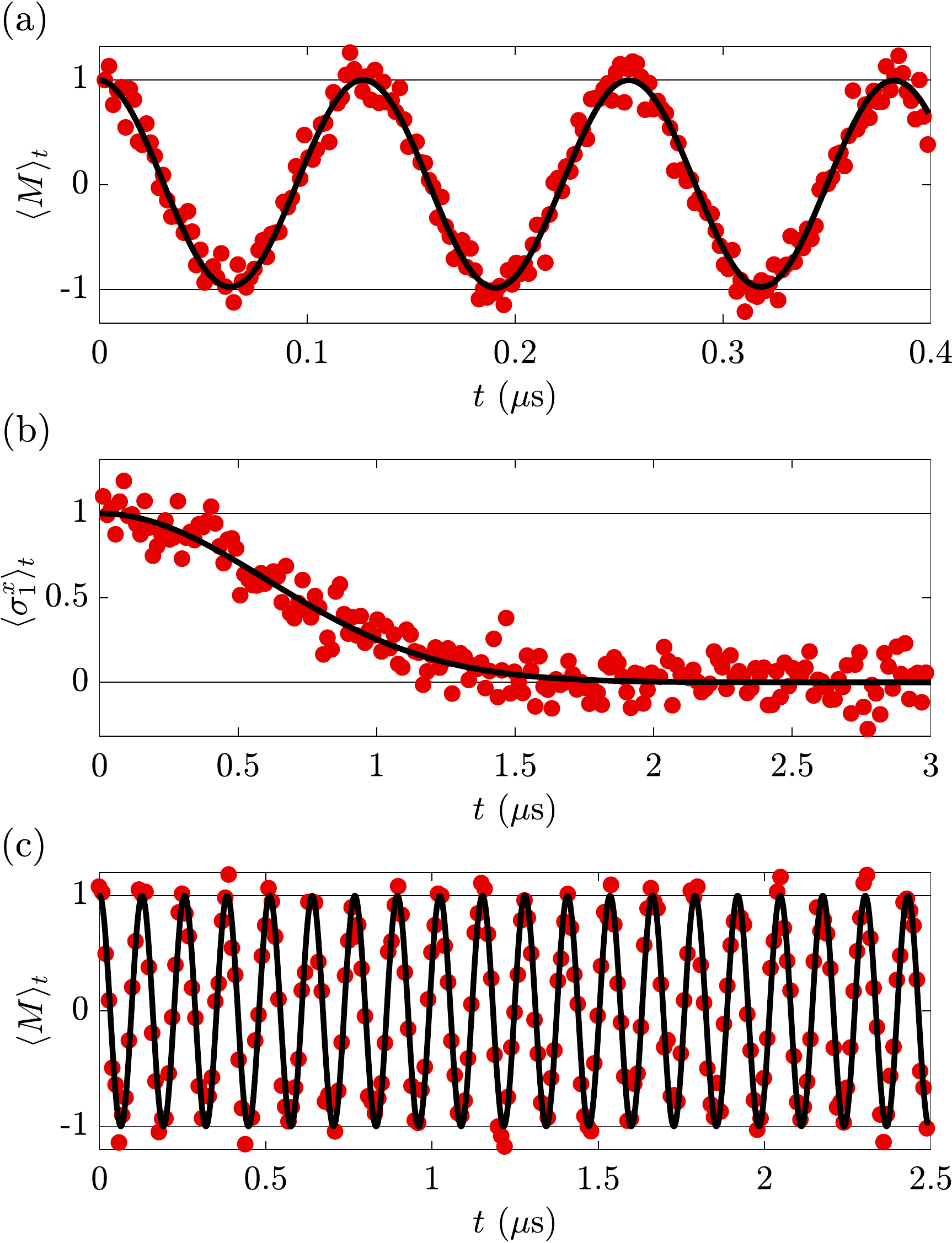}
	\caption{\label{fig6}(a) Rabi oscillations of the electron spin, i.e., driving with a constant microwave amplitude [$f(t)\equiv 1$], to determine the control field amplitude $\Omega_1/2\pi=7.91$~MHz. (b) Free induction decay experiment to determine the coherence time $T_2^\ast=0.86~\mu$s. (c) Extended Rabi oscillations to verify that the coherence time under microwave drive exceeds 2.5~$\mu$s.}  
\end{figure}

We calibrate the amplitude of the microwave driving field by measuring the frequency of Rabi oscillations of the electron spin, shown in Fig.~\ref{fig6}(a), for the given setting of the AWG and the microwave amplifier. Furthermore, we determine the coherence time $T^*_2$ of the electron spin by performing a free induction decay (FID) experiment, see Fig.~\ref{fig6}(b). Here, the FID is fitted (black line) according to $\langle\sigma_1^x\rangle_t=\exp\{-(t/T^*_2)^2\}$, yielding $T^*_2=0.86~\mu$s. However, under microwave driving the coherence time of the electron spin is significantly prolonged, as compared with the FID coherence time $T^*_2$. This is shown in Fig.~\ref{fig6}(c), where we performed a prolonged Rabi experiment. The data show no appreciable decay of the Rabi oscillations over a time span of 2.5~$\mu$s, which is the longest evolution time in our experiments. This indicates that under the microwave drive the system stays coherent sufficiently long for both the state preparation and the subsequent tomography.


\end{document}